\documentclass[sigconf,nonacm,preprint]{acmart}

\usepackage{multirow}
\usepackage{enumitem}
\usepackage{makecell}

\usepackage{algorithm}
\usepackage{algorithmic}

\usepackage{amsmath}
\usepackage{amsfonts}
\usepackage{xcolor}
\usepackage[normalem]{ulem}

\usepackage{adjustbox}
\usepackage{verbatimbox}

\AtBeginDocument{%
  }

\begin{document}

%%
%% The "title" command has an optional parameter,
%% allowing the author to define a "short title" to be used in page headers.
\title{Towards Self-cognitive Exploration: Metacognitive Knowledge Graph Retrieval Augmented Generation}

%%
%% The "author" command and its associated commands are used to define
%% the authors and their affiliations.
%% Of note is the shared affiliation of the first two authors, and the
%% "authornote" and "authornotemark" commands
%% used to denote shared contribution to the research.
% \author{Ben Trovato}
% % \authornote{Both authors contributed equally to this research.}
% \email{trovato@corporation.com}
% \orcid{1234-5678-9012}
% \author{G.K.M. Tobin}
% \authornotemark[1]
% \email{webmaster@marysville-ohio.com}
% \affiliation{%
%   \institution{Institute for Clarity in Documentation}
%   \city{Dublin}
%   \state{Ohio}
%   \country{USA}
% }

% \author{Lars Th{\o}rv{\"a}ld}
% \affiliation{%
%   \institution{The Th{\o}rv{\"a}ld Group}
%   \city{Hekla}
%   \country{Iceland}}
% \email{larst@affiliation.org}

\author{Xujie Yuan}
\affiliation{%
  \institution{Sun Yat-sen University}
  \city{Zhuhai}
  \country{China}
}
% \email{yuanxj8@mail2.sysu.edu.cn}

\author{Shimin Di}
\affiliation{%
  \institution{Southeast University}
  \city{Nanjing}
  \country{China}
}

\author{Jielong Tang}
\affiliation{%
  \institution{Sun Yat-sen University}
  \city{Zhuhai}
  \country{China}
}

\author{Libin Zheng}
\affiliation{%
  \institution{Sun Yat-sen University}
  \city{Zhuhai}
  \country{China}
}

\author{Jian Yin}
\affiliation{%
  \institution{Sun Yat-sen University}
  \city{Zhuhai}
  \country{China}
}

% \author{Aparna Patel}
% \affiliation{%
%  \institution{Rajiv Gandhi University}
%  \city{Doimukh}
%  \state{Arunachal Pradesh}
%  \country{India}}

% \author{Huifen Chan}
% \affiliation{%
%   \institution{Tsinghua University}
%   \city{Haidian Qu}
%   \state{Beijing Shi}
%   \country{China}}

% \author{Charles Palmer}
% \affiliation{%
%   \institution{Palmer Research Laboratories}
%   \city{San Antonio}
%   \state{Texas}
%   \country{USA}}
% \email{cpalmer@prl.com}

% \author{John Smith}
% \affiliation{%
%   \institution{The Th{\o}rv{\"a}ld Group}
%   \city{Hekla}
%   \country{Iceland}}
% \email{jsmith@affiliation.org}

% \author{Julius P. Kumquat}
% \affiliation{%
%   \institution{The Kumquat Consortium}
%   \city{New York}
%   \country{USA}}
% \email{jpkumquat@consortium.net}

% \author{Xujie Yuan, Shimin Di, Jielong Tang, Libin Zheng, Jian Yin}

%%
%% By default, the full list of authors will be used in the page
%% headers. Often, this list is too long, and will overlap
%% other information printed in the page headers. This command allows
%% the author to define a more concise list
%% of authors' names for this purpose.
\renewcommand{\shortauthors}{Xujie et al.}

\begin{abstract}
    Knowledge Graph-based Retrieval-Augmented Generation (KG-RAG) significantly enhances the reasoning capabilities of Large Language Models by leveraging structured knowledge. However, existing KG-RAG frameworks typically operate as open-loop systems, suffering from cognitive blindness, an inability to recognize their exploration deficiencies. This leads to relevance drift and incomplete evidence, which existing self-refinement methods, designed for unstructured text-based RAG, cannot effectively resolve due to the path-dependent nature of graph exploration. To address this challenge, We propose Metacognitive Knowledge Graph Retrieval Augmented Generation (\textbf{MetaKGRAG}), a novel framework inspired by human metacognition process, which introduces a Perceive-Evaluate-Adjust cycle to enable path-aware, closed-loop refinement. This cycle empowers the system to self-assess exploration quality, identify deficiencies in coverage or relevance, and perform trajectory-connected corrections from precise pivot points. Extensive experiments across five datasets in the medical, legal, and commonsense reasoning domains demonstrate that MetaKGRAG consistently outperforms strong KG-RAG and self-refinement baselines. Our results validates the superiority of our approach and highlights the critical need for path-aware refinement in structured knowledge retrieval.
    % we propose MetaKGRAG, a novel framework inspired by human metacognition. MetaKGRAG introduces a Perceive-Evaluate-Adjust cycle, transforming path generation into a reflective, closed-loop process that allows the model to diagnose path-level deficiencies and perform targeted, trajectory-connected corrections. We conduct extensive experiments on five datasets across the commonsense, medical, and legal domains. The results demonstrate that MetaKGRAG consistently and substantially outperforms strong baselines. This validates the superiority of our approach and highlights the critical need for path-aware refinement in structured knowledge retrieval.
\end{abstract}

%%
%% The code below is generated by the tool at http://dl.acm.org/ccs.cfm.
%% Please copy and paste the code instead of the example below.
%%
% \begin{CCSXML}
% <ccs2012>
%  <concept>
%   <concept_id>00000000.0000000.0000000</concept_id>
%   <concept_desc>Do Not Use This Code, Generate the Correct Terms for Your Paper</concept_desc>
%   <concept_significance>500</concept_significance>
%  </concept>
%  <concept>
%   <concept_id>00000000.00000000.00000000</concept_id>
%   <concept_desc>Do Not Use This Code, Generate the Correct Terms for Your Paper</concept_desc>
%   <concept_significance>300</concept_significance>
%  </concept>
%  <concept>
%   <concept_id>00000000.00000000.00000000</concept_id>
%   <concept_desc>Do Not Use This Code, Generate the Correct Terms for Your Paper</concept_desc>
%   <concept_significance>100</concept_significance>
%  </concept>
%  <concept>
%   <concept_id>00000000.00000000.00000000</concept_id>
%   <concept_desc>Do Not Use This Code, Generate the Correct Terms for Your Paper</concept_desc>
%   <concept_significance>100</concept_significance>
%  </concept>
% </ccs2012>
% \end{CCSXML}

% \ccsdesc[500]{Do Not Use This Code~Generate the Correct Terms for Your Paper}
% \ccsdesc[300]{Do Not Use This Code~Generate the Correct Terms for Your Paper}
% \ccsdesc{Do Not Use This Code~Generate the Correct Terms for Your Paper}
% \ccsdesc[100]{Do Not Use This Code~Generate the Correct Terms for Your Paper}

%%
%% Keywords. The author(s) should pick words that accurately describe
%% the work being presented. Separate the keywords with commas.
%% A "teaser" image appears between the author and affiliation
%% information and the body of the document, and typically spans the
%% page.
% \begin{teaserfigure}
%   \includegraphics[width=\textwidth]{sampleteaser}
%   \caption{Seattle Mariners at Spring Training, 2010.}
%   \Description{Enjoying the baseball game from the third-base
%   seats. Ichiro Suzuki preparing to bat.}
%   \label{fig:teaser}
% \end{teaserfigure}

% \received{20 February 2007}
% \received[revised]{12 March 2009}
% \received[accepted]{5 June 2009}

%%
%% This command processes the author and affiliation and title
%% information and builds the first part of the formatted document.
\maketitle

\section{Introduction}

\begin{figure}[t]
  \centering
  \includegraphics[width=1.0\linewidth]{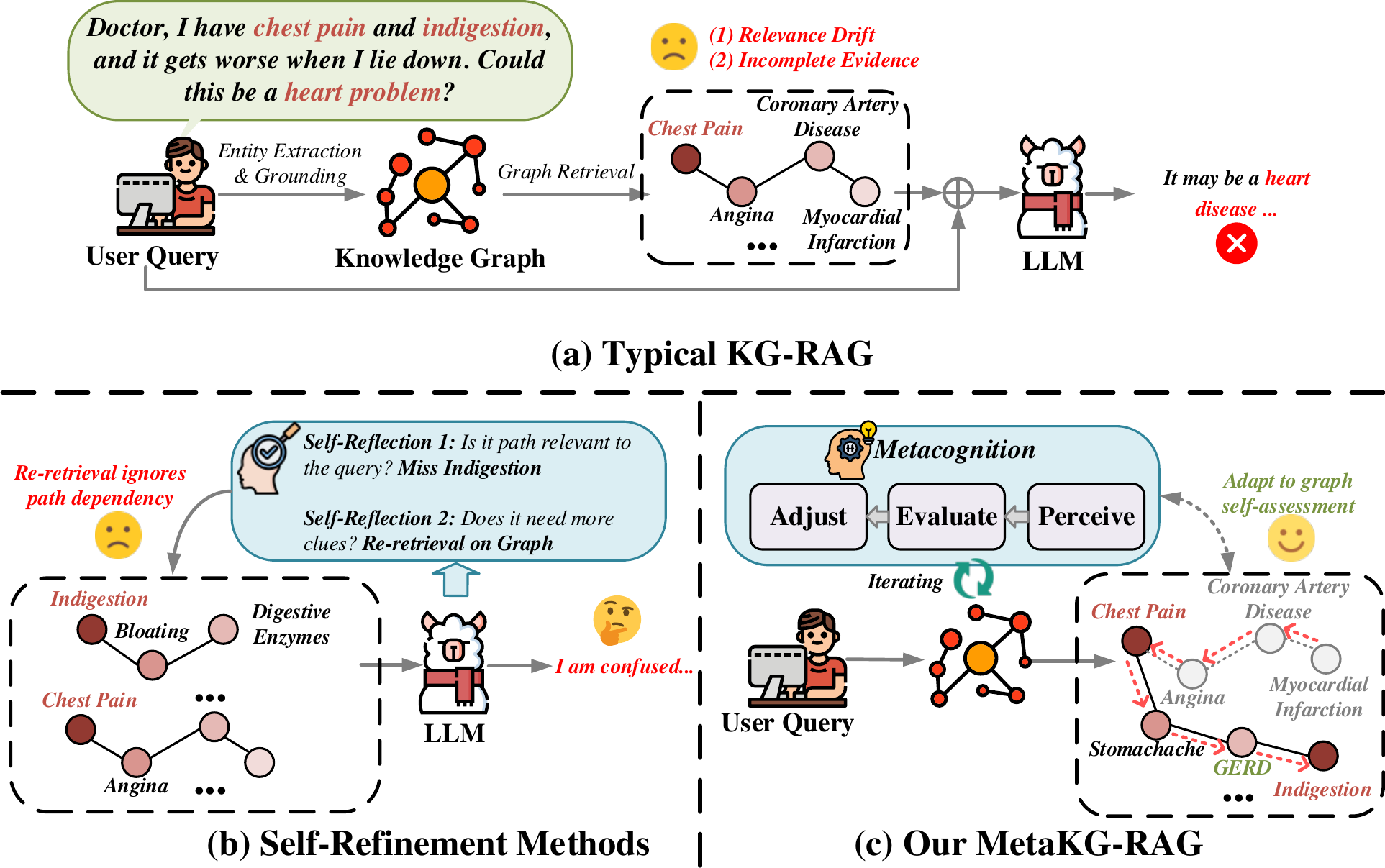}
  \caption{The comparisons of KG-RAG pipeline, Self-Refinement methods and our \textbf{MetaKGRAG}. (a) Typical KG-RAG suffers from cognitive blindness issues, leading to relevance drift and incomplete evidence. (b) Self-Refinement struggles to adapt to KG-RAG due to overlooking path dependency of graph exploration. (c) Our MetaKGRAG achieves graph-based self-cognition through a metacognitive cycle.}
  \label{fig:intro}
\end{figure}

% \footnote{
%  \# shimin: please polish the GPT style...
% }
Large Language Models (LLMs) have demonstrated remarkable capabilities~\cite{wei2022emergent,10.5555/3495724.3495883}, yet their reliability is often limited by hallucinations and outdated internal knowledge~\cite{sahoo-etal-2024-comprehensive,mallen-etal-2023-trust,Zhou2024larger_and_more}. Retrieval-Augmented Generation (RAG) mitigates these issues by grounding LLMs in external knowledge~\cite{10.5555/3524938.3525306,10.1145/3637528.3671470}. While standard RAG uses unstructured text, Knowledge Graph RAG (KG-RAG)~\cite{Edge2024FromLT,zhang-etal-2022-drlk,10.1609/aaai.v38i16.29770} leverages explicit, structured relationships. This enables a more verifiable reasoning process, making it particularly powerful for complex queries that require connecting multiple pieces of information (i.e., multi-hop reasoning) to deliver precise answers~\cite{liu-etal-2024-lost,li2025structrag,kim-etal-2023-kg,saleh-etal-2024-sg}.

Despite these advantages, the effectiveness of KG-RAG is often undermined by a challenge inherent in its exploration process. Current KG-RAG methods typically operate as \textbf{open-loop systems}, generating an evidence path in a single linear pass without a feedback mechanism~\cite{Sun2023ThinkonGraphDA, wen-etal-2024-mindmap}. This design flaw leads to what we term \textbf{Cognitive Blindness}, a state where the system is unaware of its own exploration deficiencies.
% \footnote{
%   solved \# shimin: it is a open-loop system?
% }
% \footnote{
%   solved \# shimin: please express ur idea more clearly
% }
% \footnote{
%   solved \# shimin: better introduce how the way is unaware of its own exploration deficiencies
% }
As shown in Fig.~\ref{fig:intro} (a), a real-world medical query: \textit{``Doctor, I have chest pain and indigestion, and it gets worse when I lie down. Could this be a heart problem?''}. Traditional KG-RAG methods typically employ a greedy search based on local similarity. In this case, ``chest pain'' has a strong semantic link to heart disease, biasing the system towards this path. As the exploration deepens from ``chest pain'' to ``Angina'' and then ``Coronary Artery Disease'', it progressively diverges from other crucial evidence paths, leading to \textbf{relevance drift}. 
%Jielong修改部分
As the system delves into the cardiac subgraph, this drift causes it to overlook key concepts such as ``indigestion'' and ``worse when lying down'', resulting in \textbf{incomplete evidence}. Ultimately, an incorrect answer (heart disease) is generated. Due to the absence of an effective feedback mechanism, current KG-RAG methods struggle to recognize these drifts and omissions, thereby impacting their overall performance. 

% Jielong删除部分：A consequence of this drift is \textbf{incomplete evidence}, because the system is tunneled into the cardiac subgraph, it overlooks other key concepts such as ``indigestion'' and ``worse when lying down''. 
% Since KG-RAG operates as an open-loop system without a feedback mechanism, it remains blind to its drift and omissions, potentially leading to an incorrect conclusion (heart disease) when the combined symptoms strongly suggest Gastroesophageal Reflux Disease (GERD).
% \footnote{
%   solved \# shimin: this part is important, better to explain more, and show examples
% }

% Jielong修改部分：
To address the cognitive blindness issue, a natural inclination is to apply self-refinement mechanism of current text-based RAG methods~\cite{yao2023react,jiang-etal-2023-active}. They assess discrete units of evidence (e.g., text chunks), if a flaw is found, they can substitute the faulty piece or trigger a second-time searching to fetch a new evidence. However, this paradigm is inappropriate for adapting to KG-RAG due to the \textbf{path-dependent} nature of graph exploration. The core limitation is that existing self-refinement methods treat evidence paths as a set of independent items, failing to grasp the relational trajectories between them. \textit{This is akin to realizing you are on the wrong highway, the solution is not just finding the correct road segment, but also identifying where the wrong turn occurred and re-planning a new route from your current location to the destination.} As illustrated in Fig.~\ref{fig:intro} (b), self-refinement methods recognize the missing concept ``indigestion'' and perform a re-retrieval. However, due to the absence of trajectories connecting the original ``chest pain'' path to the new ``indigestion'' path, LLMs still struggle to integrate these two separate evidence paths for the final diagnosis: \textit{heart problem} or \textit{digestive issue}? These potential knowledge conflicts across independent evidence paths induce hallucinations in LLMs. Thus, directly employing self-refinement to address the cognitive blindness issue in KG-RAG remains a challenge.

Inspired by human metacognitive processes~\cite{MetacognitiveTheories, Lai2011MetacognitionAL}, the ability to ``think about thinking'', we propose Metacognitive Knowledge Graph Retrieval-Augmented Generation (\textbf{MetaKGRAG}). This framework introduces a metacognitive cycle tailored for the path-dependent nature of graph exploration. MetaKGRAG incorporates self-monitoring and self-regulation mechanisms, which are lacking in open-loop KG-RAG systems. It transforms blind path generation into a reflective and closed-loop process through a concrete Perceive-Evaluate-Adjust cycle (as shown in Fig.~\ref{fig:intro} (c)). Here is how this cycle directly addresses cognitive blindness:

%Jielong删除部分
% This state of being ``lost'' reveals a lack of self-cognition, a core deficit that metacognition, the ability to ``think about thinking'' is designed to address~\cite{MetacognitiveTheories, Lai2011MetacognitionAL}. Metacognition provides the mechanisms of self-monitoring and self-regulation that are absent in open-loop systems. This allows us to solve Cognitive Blindness.
% To implement this metacognitive principle, we propose Metacognitive Knowledge Graph RAG (MetaKGRAG), a framework that designs a metacognitive cycle specifically for the structured nature of graph exploration. It transforms blind path generation into a reflective, closed-loop process through a concrete Perceive-Evaluate-Adjust cycle. Here is how this cycle directly cures Cognitive Blindness.

% \footnote{
%   \# move to related work, directly introduce motivation(like metaRAG, inspired from...) and our framework
% }

% \footnote{
%   solved \# shimin: explain why meta can solve cognitive blindness
% }
% To meet this challenge, we propose \textbf{Metacognitive Knowledge Graph RAG (MetaKGRAG)}, a framework that operationalizes a metacognitive cycle specifically for the structured nature of graph exploration. It is built upon a concrete \textbf{Perceive-Evaluate-Adjust} cycle designed to perceive and adjust the graph exploration, build the missing cognition.
% \footnote{
%   solved \# shimin: then explain how the new pipeline can solve several issues (including self-refinement)
% }
\begin{itemize}[nosep, leftmargin=*]
    \item \textbf{Perceive:} First, our method generates an initial candidate path and then holistically assesses it. It systematically checks how well the entire path, as a whole, covers all crucial aspects of the input query, creating a comprehensive initial understanding.
    \item \textbf{Evaluate:} Based on the Perceive, it then diagnoses specific, predefined problems. To combat \textit{Incomplete Evidence}, it identifies which key concepts from the query were missed. To correct \textit{Relevance Drift}, it pinpoints the exact node where the exploration began to deviate from the query's overall intent.
    \item \textbf{Adjust:} Different from existing self-refinement methods, the adjustment is not a blind re-retrieval. Based on the specific diagnosis, the system performs a \textbf{trajectory-connected correction}. It identifies the optimal pivot point on the flawed path, the last correct step before things went wrong. From there, it initiates a smarter re-exploration, now armed with the knowledge of what to avoid and what to prioritize, effectively re-routing its trajectory to the correct destination.
\end{itemize}

This cycle equips the system with the self-cognition to master path-dependent retrieval, providing a solution adapted to knowledge graphs.
% \footnote{
%  solved \# shimin: this part is high-level, please further polish
% }
% \footnote{
%  solved \# shimin: move claims of experiments to contributions
% }
The main contributions of this paper are as follows:

\begin{itemize}[nosep, leftmargin=*]
    \item We identify Cognitive Blindness, along with its manifestations of incomplete coverage and relevance drift, as a core challenge in current KG-RAG methods.
    \item We propose MetaKGRAG, a novel framework inspired by metacognitive principles, that introduces an evidence-level refinement cycle to overcome the limitations of traditional correction in path-dependent graph exploration.
    \item We design a concrete three-stage \texttt{Perceive-Evaluate-Adjust} cycle that enables iterative assessment, deficiency diagnosis, and strategic re-exploration of candidate evidence paths.
    \item We conduct comprehensive experiments across medical (ExplainCPE~\cite{li-etal-2023-explaincpe}, CMB-Exam~\cite{wang-etal-2024-cmb}, webMedQA~\cite{he2019applying}), legal (JEC-QA)~\cite{zhong2019jec}, and commonsense (CommonsenseQA)~\cite{talmor-etal-2019-commonsenseqa} domains demonstrate that MetaKGRAG achieves substantial improvements over strong LLM baselines, KG-RAG methods, and incorporated self-refinement approaches, validating the superiority of our adaptive control framework.
\end{itemize}

% The remainder of this paper is organized as follows. We first review related work, then detail the MetaKGRAG framework, followed by the experimental setup and results, and finally, we discuss our findings and conclude.

\section{Inspiration from Metacognition}
Metacognition~\cite{MetacognitiveTheories, Lai2011MetacognitionAL}, inspired by human "thinking about thinking" processes, has shown promise in improving system self-awareness and adaptive control. In the context of LLMs, this has led to several promising approaches. Metacognitive prompting~\cite{wang-zhao-2024-metacognitive} guides LLMs through explicit self-reflection steps, asking models to evaluate their own reasoning quality and identify potential errors.
A metacognition framework used in text-based RAG is MetaRAG~\cite{zhou2024metacognitive}, which employs a monitor-evaluate-plan loop to diagnose and rectify specific failures. It assesses an initial response by analyzing the sufficiency and consistency of both its internal knowledge and externally retrieved documents. Based on this diagnosis, it plans a targeted revision, such as generating new search queries to fill knowledge gaps or ignoring distracting evidence.

While powerful, these methods reveal a critical limitation for graph exploration that their refinement loops operate are not at the path level. For instance, MetaRAG's ``plan'' step can decide to trigger a completely new search with a revised query, but it lacks the fine-grained control to intervene within a single, ongoing graph traversal. It cannot perform fine-grained interventions, such as backtracking to a specific node on a flawed path and re-exploring from that point. Its mechanism is about replacing or adding entire evidence, not correcting a trajectory.

This highlights an adaptation gap that existing metacognitive frameworks lack the graph-native ability to perform fine-grained diagnosis and correction. They cannot answer the crucial questions, ``\textit{Which specific step in my current path was the wrong turn?}'' and ``\textit{How can I correct my trajectory from that point?}''. This gap motivates our development of MetaKGRAG, which implements metacognitive cycles specifically for the unique challenges of structured knowledge retrieval, enabling path-aware refinement.

\section{MetaKGRAG Framework}

Given a knowledge graph $\mathcal{G} = (\mathcal{E}, \mathcal{R})$ and a natural language question $Q$, the KG-RAG task is to explore $\mathcal{G}$ to retrieve a relevant evidence subgraph $\mathcal{S}$ that provides sufficient context for an LLM to generate an accurate answer. The core of this process is the generation of one or more evidence paths $P = \{(h_1, r_1, t_1), \dots, (h_n, r_n, t_n)\}$, where each triplet $(h, r, t)$ represents a head entity, relation, and tail entity respectively in the KG.

\begin{figure*}[h]
  \centering
  \includegraphics[width=0.95\linewidth]{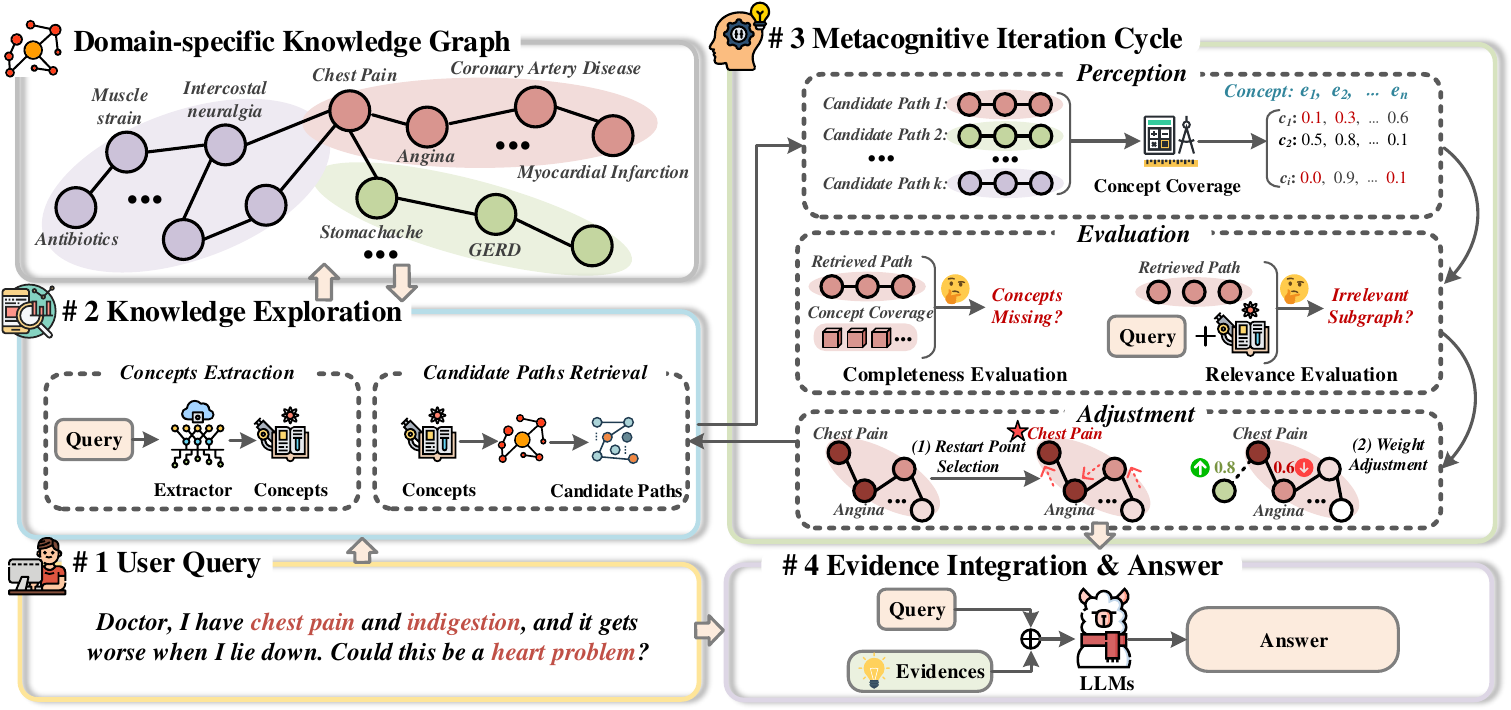}
  \caption{An overview of our MetaKGRAG framework. It iteratively refines graph exploration via a path-aware Perceive-Evaluate-Adjust cycle to address cognitive blindness in relevance drift and incomplete evidence.}
  \label{fig:model_fram}
\end{figure*}

\subsection{Framework Overview}
To simultaneously address the issues of relevance drift and incomplete evidence in a way that respects the core challenge of path dependency that simple self-refinement methods cannot solve, we designed the MetaKGRAG framework. This transforms the path generation process from a blind execution into a reflective one by operationalizing metacognitive principles into an evidence-level refinement cycle. At its core, MetaKGRAG is driven by a \textbf{Perceive-Evaluate-Adjust} cycle. This cycle is an iterative process that assesses a fully generated candidate path to inform a more strategic subsequent search. 
\begin{itemize}[nosep, leftmargin=*]
    \item \textbf{Perceive: What is the quality of my current path?} After generating an initial candidate path, the framework first Perceives its overall quality. This involves a holistic assessment of the entire path, checking how well its constituent entities cover the key concepts of the query. This step provides the raw self-awareness that traditional open-loop KG-RAG methods lack.
    \item \textbf{Evaluate: What are the specific problems?} Based on the perceived metrics, the framework Evaluates the path to diagnose specific, pre-defined issues. It explicitly checks if the path suffers from a \textit{Completeness Deficiency} (i.e., key concepts are missed) or a \textit{Relevance Deficiency} (i.e., the path has drifted into an irrelevant subgraph). This turns a vague sense of ``low quality" into an actionable diagnosis.
    \item \textbf{Adjust: How can I correct the trajectory?} In stark contrast to the disconnected re-retrievals of other methods, the adjustment is a trajectory-connected correction. Based on the diagnosis, the system performs a Strategic Re-exploration. This involves identifying the optimal pivot point on the flawed path, formulating a new search strategy, and initiating a new, smarter exploration from that point to effectively re-route the trajectory.
\end{itemize}

This iterative cycle operationalizes the principle of self-correction. By assessing a fully generated candidate path and using that diagnosis to inform a subsequent, more targeted search, the framework systematically improves the quality of evidence before it is passed to the LLM. The overall algorithm is detailed in Algorithm~\ref{alg:metakgrag}.

\begin{algorithm}[!h]
    \caption{Metacognitive Knowledge Graph RAG}
    \label{alg:metakgrag}
    \renewcommand{\algorithmicrequire}{\textbf{Input:}}
    \renewcommand{\algorithmicensure}{\textbf{Output:}}
    
    \begin{algorithmic}[1]
        \REQUIRE Question $Q$, Knowledge Graph $\mathcal{G} = (\mathcal{E}, \mathcal{R})$, Parameters $\Theta$
        \ENSURE Answer $A$
        
        \STATE $C \gets \text{ExtractConcepts}(Q, \Theta.\text{max\_concepts})$
        \STATE $E_0 \gets \text{MatchEntities}(C, \mathcal{E}, \tau_{entity})$
        \STATE $\mathcal{P}_{all} \gets \emptyset$ \hfill $\triangleright$ Collection of all refined paths 
        \FOR{each entity $e \in E_0$}
            \STATE $P_{candidate} \gets \text{InitialPathSearch}(e, Q, \mathcal{G})$ 
            % \hfill $\triangleright$ Generate an initial candidate path
            \FOR{$i = 1$ to $N_{max}$}
                \STATE $\mathcal{M} \gets \text{Perceive}(P_{candidate}, C)$ \hfill $\triangleright$ Assess overall quality 
                \STATE $\mathcal{I} \gets \text{Evaluate}(\mathcal{M}, Q, C, \Theta)$ \hfill $\triangleright$ Diagnose deficiencies
                \IF{$\mathcal{I} = \emptyset$}
                    \STATE \textbf{break} \hfill $\triangleright$ No issues found
                \ENDIF
                \STATE $P_{new} \gets \text{Adjust}(P_{candidate}, \mathcal{I}, Q, \mathcal{G})$ \hfill $\triangleright$ Initiate a new search to generate a refined path
                \IF{$\text{PathSimilarity}(P_{new}, P_{candidate}) > \tau_{similarity}$}
                    \STATE \textbf{break} \hfill $\triangleright$ No significant improvement
                \ENDIF
                \STATE $P_{candidate} \gets P_{new}$
            \ENDFOR
            \IF{$P_{candidate} \neq \emptyset$}
                \STATE $\mathcal{P}_{all} \gets \mathcal{P}_{all} \cup \{P_{candidate}\}$ \hfill 
            \ENDIF
        \ENDFOR
        
        \STATE $\mathcal{S} \gets \text{IntegrateEvidencePaths}(\mathcal{P}_{all})$
        \STATE $A \gets \text{GenerateAnswer}(Q, \mathcal{S})$
        \STATE \textbf{return} $A$
    \end{algorithmic}
\end{algorithm}

\subsection{Initial Knowledge Exploration}
Before entering the metacognitive cycle, MetaKGRAG establishes multiple starting points to ensure comprehensive exploration.
Given question $Q$, we use an LLM to extract key concepts $C = \{c_1, c_2, ..., c_k\}$ that represent the essential information needs. This extraction is domain-aware, prioritizing technical terms and named entities relevant to the task. For each concept $c_i$, we identify corresponding entities in $\mathcal{G}$ through semantic matching. Specifically, we compute the semantic similarity:
$$\text{sim}(c_i, e_j) = \frac{\mathbf{v}_{c_i} \cdot \mathbf{v}_{e_j}}{||\mathbf{v}_{c_i}|| \cdot ||\mathbf{v}_{e_j}||}$$
where $\mathbf{v}_{c_i}$ and $\mathbf{v}_{e_j}$ are the embedding vectors of concept $c_i$ and entity $e_j$ respectively.
To ensure quality, we only select the most semantically similar entities for each concept, subject to a minimum similarity threshold:
$$E_0 = \{\arg\max_{e \in \mathcal{E}} \text{sim}(c_i, e) : c_i \in C\}$$
where we only retain entities with $\text{sim}(c_i, e) > \tau_{entity}$ to ensure high-quality matches, and $\tau_{entity}$ is the entity matching threshold.
This results in a set of high-confidence starting points $E_0 = \{e_1, e_2, ..., e_m\}$. Each starting entity initiates an independent retrieval path. This multi-perspective approach ensures that different aspects of the query are explored, reducing the risk of missing crucial information due to a single suboptimal starting point. 
For each starting entity $e \in E_0$, we first generate an initial candidate path $P_{candidate}$ through greedy exploration from $e$, following edges with the highest relevance to query $Q$.

\subsection{Metacognitive Iteration Cycle}
This iterative cycle is the core of our framework. It takes a candidate evidence path and progressively refines it by diagnosing and correcting its flaws. The cycle consists of three distinct modules: Perception, Evaluation, and Adjustment.

\begin{itemize}[nosep, leftmargin=*]
    \item \textbf{Perception Module.} The Perception module acts as a quality sensor for the initial candidate path. Let $E_P = \{e : e \in \text{head or tail of any triple in } P_{candidate}\}$ be the set of entities in the path. It computes a coverage map $\mathcal{M}$ by assessing how well the path covers each key concept $c_i \in C$:
    $$\text{Coverage}(c_i) = \max_{e \in E_P} \text{sim}(c_i, e)$$
    This provides the raw data for self-assessment.

    \item \textbf{Evaluation Module.} Based on the output from the Perception module, the Evaluation module acts as a diagnostic engine. It formalizes and diagnoses the two key deficiencies discussed in the introduction. A path $P$ may suffer from:
    \begin{itemize}
        \item \textbf{Completeness Deficiency,} if it fails to cover all key aspects. Formally, a path $P$ is deficient in completeness if there exists a concept $c_i \in C$ for which the coverage, defined as $\max_{e \in \text{entities}(P)} \text{sim}(c_i, e)$, falls below a threshold $\tau_{coverage}$.
        \item \textbf{Relevance Deficiency,} if it contains entities that are only locally relevant to one aspect but have low relevance to the overall query $Q$. Formally, a path $P$ is deficient in relevance if it contains entities with a low $\text{GlobalSupport}$ score.
    \end{itemize}
    To implement this, the module analyzes the path entities to identify specific deficiencies:
    \begin{enumerate}
        \item \textit{To detect Completeness Deficiency,} it flags concepts whose coverage scores are below a threshold $\tau_{coverage}$ as the set of missing concepts $\mathcal{C}_{missing}$. This indicates that the current path has failed to explore entities relevant to certain key concepts.
        \item \textit{To detect Relevance Deficiency,} it scrutinizes each entity in the path by calculating its $\text{GlobalSupport}(e, Q, C)$. This metric balances concept coverage with overall question relevance:
        $$\text{EntityScope}(e, C) = \frac{|\{c \in C : \text{sim}(e, c) > \tau_c\}|}{|C|}$$
        $$\text{GlobalSupport}(e, Q, C) = \alpha \cdot \text{EntityScope}(e, C) + (1-\alpha) \cdot \text{sim}(e, Q)$$
        where $\text{EntityScope}(e, C)$ measures how many concepts entity $e$ is relevant to, $\alpha$ balances the two components, and $\tau_c$ is the concept relevance threshold. Entities with low global support are flagged as misleading.
    \end{enumerate}
    The evaluation module produces a diagnosis $\mathcal{I}$ of the issues found in $P_{candidate}$.

    \item \textbf{Adjustment Module.} If deficiencies are diagnosed ($\mathcal{I} \neq \emptyset$), this module executes a corrective action. Instead of triggering new, independent searches, our approach performs a targeted correction of the flawed trajectory itself. This Strategic re-exploration process involves: 
    \begin{enumerate}
        \item \textit{Formulating an Adjustment Strategy}. For a Completeness Deficiency, the system identifies external entities relevant to the missing concepts $\mathcal{C}_{missing}$ and assigns them a positive weight adjustment $+\delta$. For a Relevance Deficiency, it assigns a negative weight adjustment $-\delta$ to misleading entities. These adjustments modify the original edge weights during subsequent exploration:
        $$w_{adjusted}(e_i, e_j) = w_{original}(e_i, e_j) + \text{adjustment}(e_j)$$
        \item \textit{Executing an Informed Re-search}. Rather than restarting from scratch, the module selects a strategic restart point from the previous path. For Completeness Deficiency, it chooses the entity most relevant to the missing concepts. For Relevance Deficiency, it selects the entity with the highest global support:
        $$e_{restart} = \arg\max_{e \in E_P} f(e)$$
        where 
        $$f(e) = \begin{cases}
        \max_{c \in \mathcal{C}_{missing}} \text{sim}(e, c) & \text{Completeness issue} \\
        \text{GlobalSupport}(e, Q, C) & \text{ Relevance issue}
        \end{cases}$$
        The system then performs greedy search from $e_{restart}$, selecting the next entity with maximum adjusted weight at each step until a new candidate path is generated.
    \end{enumerate}
    This targeted re-exploration mechanism directly addresses the diagnosed issues, efficiently correcting the path's trajectory without losing all prior progress.
\end{itemize}

The metacognitive cycle repeats until one of three convergence conditions is met: (1) the Evaluation module detects no deficiencies, (2) the change between the new path and the previous one is minimal, measured by entity overlap similarity exceeding a threshold $\tau_{similarity}$, or (3) a maximum iteration limit $N_{max}$ is reached. Specifically, two paths are considered similar if:
$$\text{PathSimilarity}(P_1, P_2) = \frac{|E_{P_1} \cap E_{P_2}|}{|E_{P_1} \cup E_{P_2}|} > \tau_{similarity}$$
where $E_{P_1}$ and $E_{P_2}$ are the entity sets of the two paths. In our implementation, we set the default $N_{max} = 3$ and $\tau_{similarity} = 0.8$ based on empirical validation.

\subsection{Multi-Evidence Integration and Answer Generation}
After all initial entities complete their independent metacognitive cycles, MetaKGRAG performs comprehensive multi-path integration to synthesize the collected evidence. The evidence synthesis process merges all refined paths into a comprehensive evidence subgraph $\mathcal{S}$ while removing duplicate information, thus preserving diverse perspectives while eliminating redundancy. 

The evidence subgraph is then converted to natural language context through a structured transformation. Each triple $(h_i, r_i, t_i) \in \mathcal{S}$ is converted into natural language statements following the template ``Evidence $i$: $h_i$ $r_i$ $t_i$'', creating a structured evidence list that maintains the semantic relationships between entities while being interpretable by the LLM. Finally, the LLM generates the answer $A$ using both the original question $Q$ and the curated evidence context through a prompt that instructs the model to analyze the evidence and answer the question. The complete prompt template for answer generation is detailed in Appendix~\ref{app:answer_prompt}.

\section{Experiments}

\subsection{Experimental Setup}
\subsubsection{Datasets}
As shown in Tab~\ref{tab:dataset_statistics}, to evaluate MetaKGRAG's effectiveness and generalization ability across diverse domains, we conduct experiments on five datasets spanning commonsense, medical, and legal domains. 

\begin{table}[h]
    \centering
    \caption{The statistics of datasets.}
    \label{tab:dataset_statistics}
    \begin{tabular}{llcl}
    \hline
    \textbf{Domain} & \textbf{Dataset} & \textbf{Questions} & \textbf{Language} \\ \hline
    Commonsense & CommonsenseQA & 700 & English \\ \hline
    \multirow{3}{*}{Medical} & CMB-Exam & 2,000 & Chinese \\
     & ExplainCPE & 507 & Chinese \\
     & webMedQA & 500 & Chinese \\ \hline
    Legal & JEC-QA & 479 & Chinese \\ \hline
    \end{tabular}
\end{table}

We use CommonsenseQA~\cite{talmor-etal-2019-commonsenseqa} for commonsense knowledge evaluation, which contains multiple-choice questions requiring broad knowledge integration. For medical domains, we employ three datasets, CMB-Exam~\cite{wang-etal-2024-cmb} with 2,000 sampled questions from Chinese medical professional examinations (including Nursing, Pharmacy, Postgraduate, and  Professional), ExplainCPE~\cite{li-etal-2023-explaincpe} containing pharmaceutical questions from the National Licensed Pharmacist Examination with both answers and explanations, and webMedQA~\cite{he2019applying} featuring real-world patient-doctor conversations from online medical platforms. For legal domain, we use JEC-QA~\cite{zhong2019jec} from China's National Judicial Examination, which requires logical reasoning to apply legal materials to specific case scenarios.
For more detailed information on these datasets, please refer to Appendix~\ref{appx:exp-data}.

\subsubsection{Evaluation Metrics}
For evaluation, we adopt a variety of different metrics. \textit{Correct (Accuracy)}, \textit{Wrong}, \textit{Fail} are used for those with ground truth (e.g., CommonsenseQA, CMB-Exam, ExplainCPE), where \textit{Fail} indicates the model fails to generate any answer.  For tasks requiring generative answers (ExplainCPE, webMedQA), we use \textit{ROUGE-L}~\cite{lin-2004-rouge} to measure lexical overlap with reference answers and \textit{BERTScore}~\cite{bert-score} to assess semantic similarity. To further evaluate the overall quality of the generated responses, we utilize \textit{G-Eval}~\cite{liu-etal-2023-g}, a framework that leverages LLMs for evaluation, assessing the generated answers based on four key dimensions, \textit{Coherence}, \textit{Consistency}, \textit{Fluency}, and \textit{Relevance}. For all results, the best results are in \textbf{bold} and the second best results are \underline{underlined}.

\subsubsection{Baselines}
To comprehensively evaluate the performance of our proposed MetaKGRAG framework, we select a diverse range of baselines, which can be categorized into three groups: Large Language Models, KG-RAG approaches, and self-refinement methods.

\begin{itemize}[nosep, leftmargin=*]
    \item \textit{Large Language Models.} This group serves as a fundamental baseline to evaluate the capabilities of LLMs themselves without any external knowledge retrieval. We select leading commercial models including \textbf{GPT-4o}, \textbf{Claude 3.5 Sonnet}, \textbf{Gemini 1.5 Pro}, and OpenAI \textbf{o1-mini}. For open-source models, we use \textbf{Qwen2.5-7B} and \textbf{Qwen2.5-72B}~\cite{yang2024qwen2.5} for Chinese tasks, and \textbf{Llama-3-8B} and \textbf{Llama-3-70B}~\cite{dubey2024llama} for English tasks. 
    \item \textit{KG-RAG Approaches.} We compare against several representative KG-RAG approaches. \textbf{Vanilla KGRAG}~\cite{Soman2023BiomedicalKG} serves as the fundamental implementation, performing direct entity similarity-based retrieval and presenting facts to LLMs. \textbf{ToG (Think-on-Graph)}~\cite{Sun2023ThinkonGraphDA} guides LLMs to explore multiple reasoning paths within knowledge graphs for multi-hop reasoning. \textbf{MindMap}~\cite{wen-etal-2024-mindmap} constructs structured representations that integrate knowledge from subgraphs to enhance interpretability. \textbf{KGGPT}~\cite{kim-etal-2023-kg} handles complex queries by decomposing them into simpler clauses and constructing evidence graphs through separate retrievals. These baselines providing comprehensive coverage of existing retrieval paradigms from simple similarity matching to sophisticated multi-hop strategies.
    \item \textit{Self-Refinement Methods.} As discussed in Sec.~\ref{relatedworks}, we compare several Self-Refinement methods. \textbf{Chain-of-Thought (CoT)}~\cite{wei2022chain} encourages more thoughtful reasoning by generating intermediate steps. \textbf{Metacognitive Prompting}~\cite{wang-zhao-2024-metacognitive} employs a metacognitive prompt to guide the model to self-critique and refine its reasoning. For retrieval-augmented scenarios, we construct baselines by combining a standard KG-RAG retriever with frameworks like \textbf{FLARE}~\cite{jiang-etal-2023-active}, which performs active retrieval when generation confidence is low, and \textbf{ReAct}~\cite{yao2023react}, which synergizes reasoning and acting to iteratively search for information. We also adapt \textbf{Meta RAG}~\cite{zhou2024metacognitive}, which enhances RAG by implementing a three-step metacognitive process of monitoring, evaluating, and planning to enable the model to introspectively identify and rectify its own knowledge gaps and reasoning errors. These baselines represent a straightforward ``stacking'' approach that adds a refinement mechanism in KG-RAG.
\end{itemize}

\subsubsection{Implementation Details}
Our MetaKGRAG framework is implemented using both large and small-scale open-source models as its backbone. Specifically, we utilize the Qwen2.5 series (7B, 72B) for Chinese tasks and the Llama-3 series (8B, 70B) for English tasks. For all semantic similarity calculations, such as matching concepts to entities and assessing path coverage, we employ the distiluse-base-multilingual-cased-v1~\cite{reimers-2019-sentence-bert} embedding model due to its strong multilingual capabilities. The specific knowledge graphs for each dataset were custom-built to ensure relevance and quality. The detailed methodologies for knowledge graph construction, the specific prompt templates used for different stages of the framework (e.g., concept extraction, answer generation), and the experimental environment setup are all detailed in the Appendix~\ref{appx:exp} for reproducibility. The core parameters of MetaKGRAG will be analyzed in detail in the subsequent analysis section.

\begin{table}[t] 
  \centering
  \caption{Performance Comparison on ExplainCPE and JEC-QA using Accuracy.}
  \label{tab:main_results_med_legal}
  \footnotesize
  \setlength\tabcolsep{12pt}
  \renewcommand{\arraystretch}{1.2}  
  \begin{tabular}{c|c|c|c}
    \hline
    \textbf{Type} & \textbf{Method} & \textbf{ExplainCPE} & \textbf{JEC-QA} \\
    % & & \textbf{(Correct \%)} & \textbf{(Correct \%)} \\ 
    \hline
    \multicolumn{4}{c}{\textbf{Without Retrieval}} \\ \hline
    \multirow{6}{*}{LLM Only} & Qwen2.5-7B & 69.76 & 65.06 \\ 
      & Qwen2.5-72B & 81.82 & 80.13 \\ 
      & GPT4o & 79.64 & 78.51 \\ 
      & o1-mini & 75.10 & 70.15 \\ 
      & Claude3.5-Sonnet & 76.88 & 75.33 \\ 
      & Gemini1.5-Pro & 69.37 & 76.18 \\ \hline
      \multirow{2}{*}{Self-Refine} 
      & Chain-of-Thought & 82.53 & 81.02 \\ 
      & Meta Prompting & 83.11 & \underline{81.67} \\
      \hline
    \multicolumn{4}{c}{\textbf{With Retrieval}} \\ \hline
      \multirow{4}{*}{KG-RAG} 
      & KGRAG & 78.53 & 73.88 \\ 
      & ToG & 78.85 & 74.90 \\ 
      & MindMap & 78.41 & 71.55 \\ 
      & KGGPT & 78.86 & 71.83 \\
      \hline
      \multirow{4}{*}{Self-Refine} 
      & FLARE & 80.23 & 75.81 \\ 
      & ReAct & 81.51 & 76.92 \\ 
      & Meta Prompting & 80.88 & 76.25 \\ 
      & Meta RAG & 81.93 & 77.31 \\
      \hline
      \multirow{2}{*}{Ours} 
      & \makecell{MetaKGRAG \\ (Qwen2.5-7B)} & \underline{85.97} & 77.10 \\ 
      & \makecell{MetaKGRAG \\ (Qwen2.5-72B)} & \textbf{91.70} & \textbf{88.49} \\
      \hline
  \end{tabular}
  % \vspace{-10pt}  
\end{table}

\begin{table}[htbp] 
  \centering
  \caption{Performance Comparison on webMedQA. Prec. and Rec. represent Precision and Recall, respectively.}
  \label{tab:webmedqa_results}
  \footnotesize
  \setlength\tabcolsep{8pt}
  \renewcommand{\arraystretch}{1.2}  
  \begin{tabular}{c|c|ccc}
    \hline
    \textbf{Type} & \textbf{Method} & \textbf{Prec.} & \textbf{Rec.} & \textbf{F1} \\ \hline
    \multicolumn{5}{c}{\textbf{Without Retrieval}} \\ \hline
    \multirow{6}{*}{LLM Only} & Qwen2.5-7B & 66.68 & 71.14 & 68.85 \\ 
      & Qwen2.5-72B & 70.12 & 73.58 & 71.81 \\ 
      & GPT4o & 72.53 & 75.11 & 73.80 \\ 
      & o1-mini & 68.24 & 72.05 & 70.09 \\ 
      & Claude3.5-Sonnet & 71.89 & 74.32 & 73.09 \\ 
      & Gemini1.5-Pro & 72.01 & 74.88 & 73.42 \\ \hline
      \multirow{2}{*}{Self-Refine} 
      & Chain-of-Thought & 71.21 & 74.63 & 72.88 \\ 
      & Meta Prompting & 71.85 & 75.01 & 73.40 \\
      \hline
    \multicolumn{5}{c}{\textbf{With Retrieval}} \\ \hline
      \multirow{4}{*}{KG-RAG} 
      & KGRAG & 73.15 & 76.02 & 74.56 \\ 
      & ToG & 73.89 & 76.55 & 75.19 \\ 
      & MindMap & 72.93 & 75.81 & 74.34 \\ 
      & KGGPT & 73.51 & 76.23 & 74.85 \\
      \hline
      \multirow{4}{*}{Self-Refine} 
      & FLARE & 74.22 & 76.91 & 75.54 \\ 
      & ReAct & 74.98 & 77.53 & 76.23 \\ 
      & Meta Prompting & 74.53 & 77.18 & 75.83 \\ 
      & Meta RAG & 75.11 & 77.82 & 76.44 \\
      \hline
      \multirow{2}{*}{Ours} 
      & \makecell{MetaKGRAG \\ (Qwen2.5-7B)} & \underline{76.53} & \underline{78.91} & \underline{77.70} \\ 
      & \makecell{MetaKGRAG \\ (Qwen2.5-72B)} & \textbf{78.02} & \textbf{80.15} & \textbf{79.07} \\
      \hline
  \end{tabular}
  % \vspace{-10pt}  
\end{table}

\begin{table}[htbp]
  \centering
  \caption{Performance comparison on CommonsenseQA.}
  \label{tab:main_results_csqa}
  \footnotesize
  \setlength\tabcolsep{4.5pt} % single column
  \renewcommand{\arraystretch}{1.2}  
  \begin{tabular}{l|l|ccc}
    \hline
    \textbf{Type} & \textbf{Method} & \textbf{Correct} & \textbf{Wrong} & \textbf{Fail} \\ \hline
    \multicolumn{5}{c}{\textbf{Without Retrieval}} \\ \hline
    \multirow{6}{*}{LLM Only} & Llama-3-8B & 73.82 & 26.04 & 0.14 \\ 
      & Llama-3-70B & 81.76 & 18.24 & 0.00 \\ 
      & GPT-4o & 84.54 & 15.44 & 0.02 \\ 
      & o1-mini & 81.41 & 18.45 & 0.14 \\ 
      & Claude3.5-Sonnet & 82.55 & 17.45 & 0.00 \\ 
      & Gemini1.5-Pro & 83.83 & 16.17 & 0.00 \\ \hline
    \multirow{2}{*}{Self-Refine} 
      & Chain-of-Thought & 83.52 & 16.48 & 0.00 \\ 
      & Meta Prompting & 84.13 & 15.87 & 0.00 \\
      \hline
    \multicolumn{5}{c}{\textbf{With Retrieval}} \\ \hline
    \multirow{4}{*}{KG-RAG} 
      & KGRAG & 85.04 & 14.96 & 0.00 \\ 
      & ToG & 85.81 & 14.19 & 0.00 \\ 
      & MindMap & 85.53 & 14.47 & 0.00 \\ 
      & KGGPT & 86.22 & 13.78 & 0.00 \\
      \hline
    \multirow{4}{*}{Self-Refine} 
      & FLARE & 86.54 & 13.46 & 0.00 \\ 
      & ReAct & 87.31 & 12.69 & 0.00 \\ 
      & Meta Prompting & 86.81 & 13.17 & 0.02 \\ 
      & Meta RAG & 87.52 & 12.48 & 0.00 \\
      \hline
    \multirow{2}{*}{Ours} 
      & MetaKGRAG (Llama-3-8B) & \underline{88.54} & 11.46 & 0.00 \\ 
      & MetaKGRAG (Llama-3-70B) & \textbf{92.11} & 7.89 & 0.00 \\
      \hline
  \end{tabular}
\end{table}

\begin{table*}[htbp] 
  \centering
  \caption{Performance Comparison on CMB-Exam with six different types.}
  \label{tab:main_results_cmb_exam}
  \footnotesize
  \setlength\tabcolsep{4pt}
  \renewcommand{\arraystretch}{1.2}  
  \begin{tabular}{c|c|ccc|ccc|ccc|ccc}
    \hline
    \multirow{2}{*}{\textbf{Type}} & \multirow{2}{*}{\textbf{Method}} & \multicolumn{3}{c|}{\textbf{Nursing}}  &  \multicolumn{3}{c|}{\textbf{Pharmacy}} &  \multicolumn{3}{c|}{\textbf{Postgraduate}} &  \multicolumn{3}{c}{\textbf{Professional}} \\ \cline{3-14}
     &  & \textbf{Correct} & \textbf{Wrong} & \textbf{Fail} & \textbf{Correct} & \textbf{Wrong} & \textbf{Fail} & \textbf{Correct} & \textbf{Wrong} & \textbf{Fail} & \textbf{Correct} & \textbf{Wrong} & \textbf{Fail} \\ \hline
    \multicolumn{14}{c}{\textbf{Without Retrieval}} \\ \hline
    \multirow{6}{*}{LLM Only} & Qwen2.5-7B & 80.96 & 18.84 & 0.20 & 77.56 & 22.24 & 0.20 & 80.36 & 19.64 & 0.00 & 74.15 & 25.85 & 0.00 \\ 
      & Qwen2.5-72B & 89.80 & 10.08 & 0.12 & 90.08 & 9.92 & 0.00 & 88.18 & 12.62 & 0.20 & 83.98 & 16.02 & 0.00 \\ 
      & GPT4o & 83.13 & 16.87 & 0.00 & 72.89 & 26.91 & 0.20 & 76.95 & 22.44 & 0.60 & 78.96 & 21.04 & 0.00 \\ 
      & o1-mini & 74.50 & 25.50 & 0.00 & 60.44 & 39.56 & 0.00 & 63.13 & 36.27 & 0.60 & 73.55 & 26.45 & 0.00 \\ 
      & Claude3.5-Sonnet & 75.90 & 24.10 & 0.00 & 65.86 & 34.14 & 0.00 & 69.54 & 30.46 & 0.00 & 73.75 & 26.25 & 0.00 \\ 
      & Gemini1.5-Pro & 80.72 & 19.28 & 0.00 & 70.68 & 29.32 & 0.00 & 75.95 & 24.05 & 0.00 & 77.56 & 22.44 & 0.00 \\ \hline
      \multirow{2}{*}{Self-Refine} 
      & CoT & 90.15 & 9.73 & 0.12 & 90.31 & 9.69 & 0.00 & 88.54 & 11.26 & 0.20 & 84.22 & 15.78 & 0.00 \\ 
      & Meta Prompting & 90.55 & 9.33 & 0.12 & 90.72 & 9.28 & 0.00 & 88.91 & 10.89 & 0.20 & 84.67 & 15.33 & 0.00 \\
      \hline
    \multicolumn{14}{c}{\textbf{With Retrieval}} \\ \hline
      \multirow{4}{*}{KG-RAG} 
      & KGRAG & 88.15 & 11.85 & 0.00 & 85.33 & 14.47 & 0.20 & 85.12 & 14.88 & 0.00 & 82.71 & 17.29 & 0.00 \\ 
      & ToG & 89.18 & 10.62 & 0.20 & 86.77 & 13.23 & 0.00 & 85.17 & 14.83 & 0.00 & 83.37 & 16.63 & 0.00 \\ 
      & MindMap & 85.77 & 14.02 & 0.20 & 81.95 & 17.65 & 0.41 & 80.97 & 18.83 & 0.00 & 81.19 & 18.81 & 0.00 \\ 
      & KGGPT & 86.74 & 13.26 & 0.00 & 86.13 & 13.87 & 0.00 & 86.15 & 13.85 & 0.00 & 83.62 & 16.38 & 0.00 \\ 
      \hline
      \multirow{4}{*}{Self-Refine} 
      & FLARE & 89.11 & 10.89 & 0.00 & 87.15 & 12.65 & 0.20 & 87.52 & 12.48 & 0.00 & 85.18 & 14.82 & 0.00 \\ 
      & ReAct & 90.12 & 9.88 & 0.00 & 88.23 & 11.57 & 0.20 & 88.43 & 11.57 & 0.00 & 86.27 & 13.73 & 0.00 \\ 
      & Meta Prompting & 89.55 & 10.45 & 0.00 & 87.64 & 12.16 & 0.20 & 87.91 & 12.09 & 0.00 & 85.73 & 14.27 & 0.00 \\ 
      & Meta RAG & 90.53 & 9.47 & 0.00 & 88.71 & 11.09 & 0.20 & 88.88 & 11.12 & 0.00 & 86.74 & 13.26 & 0.00 \\
      \hline
      \multirow{2}{*}{Ours} 
      & MetaKGRAG (Qwen2.5-7B) & \underline{91.78} & 8.22 & 0.00 & \underline{91.58} & 8.42 & 0.00 & \underline{89.78} & 10.22 & 0.00 & \underline{88.78} & 11.22 & 0.00 \\ 
      & MetaKGRAG (Qwen2.5-72B) & \textbf{96.54} & 3.46 & 0.00 & \textbf{98.79} & 1.21 & 0.00 & \textbf{92.99} & 7.01 & 0.00 & \textbf{94.59} & 5.41 & 0.00 \\
      \hline
  \end{tabular}
  % \vspace{-10pt}
\end{table*}

\subsection{Results and Analysis}

\subsubsection{Main Results}
The main experimental results are presented in Tab.~\ref{tab:main_results_med_legal}, \ref{tab:webmedqa_results}, \ref{tab:main_results_csqa}, and \ref{tab:main_results_cmb_exam}. The results across all five datasets validate the effectiveness of our framework. We highlight three key findings.
% \begin{enumerate}

(1) Our framework consistently outperforms all baselines, achieving 91.70\% on ExplainCPE (+9.88\% over best LLM), 92.11\% on CommonsenseQA (+10.35\%), and 88.49\% on JEC-QA (+8.36\%). These improvements across medical, commonsense, and legal domains demonstrate the effectiveness of our metacognitive approach. Regarding to the ROUGE-L and G-Eval results of webMedQA and ExplainCPE, please refer to the Appendix~\ref{appx:exp-results}. As shown in Tab.~\ref{tab:results_rouge_geval}, our MetaKGRAG method also achieved scores that surpass other baselines. The results demonstrate that our method not only improves the accuracy of multiple-choice question answering, but also excels in explanatory answer generation capabilities.

(2) A critical finding is that simply stacking existing refinement methods on KG-RAG yields only marginal improvements. Methods like FLARE, ReAct, and Meta RAG. When combined with KG-RAG, improve performance by merely 1-3\%. In stark contrast, MetaKGRAG achieves 5-10\% improvements over basic KG-RAG methods. This dramatic difference reveals that document-oriented refinement strategies fail to address the unique challenges of graph exploration. Our perceive-evaluate-adjust cycle, specifically designed to handle path dependencies and structural constraints, proves essential for high-quality evidence retrieval in KG.

(3) MetaKGRAG demonstrates effectiveness with both small and large backbone models. With smaller models (7B/8B), MetaKGRAG already achieves competitive performance, for example, reaching 85.97\% on ExplainCPE. When scaled to larger models (72B/70B), the performance gains are even more pronounced, suggesting that stronger base models can better leverage the high-quality evidence paths produced by our metacognitive cycle. 
% \end{enumerate}

\subsubsection{Ablation Studies}
To gain deeper insights into the specific mechanisms driving MetaKGRAG's performance, we conduct fine-grained ablation studies that examine individual functional components within our framework. Rather than removing entire \textbf{Metacognitive Cycle}, we selectively disable specific mechanisms to understand their individual contributions. In Evaluation module, \textbf{Completeness Check }identifies missing key concepts, \textbf{Relevance Check} detects entities to prevent misleading information. In Adjustment module, \textbf{Strategic Restart} intelligently selects optimal starting points for path re-exploration when issues are detected, rather than naively restarting from the original entity.

Table \ref{tab:ablation_fine_grained} shows that Completeness Check provides the largest contribution (5.04\% on ExplainCPE, 5.48\% on JEC-QA and 2.39\% on CommonsenseQA), indicating that ensuring comprehensive information coverage is more critical than filtering irrelevant content. Relevance Check contributes substantially (4.32\%, 4.22\%, and 1.99\% respectively), demonstrating the importance of avoiding misleading entities that satisfy keyword matching without providing meaningful information. Strategic Restart offers consistent but modest gains (3.16\%, 3.56\%, and 1.29\%), proving that intelligent restart point selection outperforms naive re-exploration strategies. 

\begin{table}[htbp]
    \centering
    \caption{Ablation study on ExplainCPE, JEC-QA, and CommonsenseQA with backbone model Qwen2.5-7B.}
    \label{tab:ablation_fine_grained}
    \footnotesize
    \setlength\tabcolsep{4.5pt}
    \begin{tabular}{lccc}
    \hline
    \textbf{Configuration} & \textbf{ExplainCPE} & \textbf{JEC-QA} & \textbf{CommonsenseQA} \\
    \hline
    MetaKGRAG (Original) & 85.97 & 77.10 & 88.54 \\
    \quad w/o Metacognitive Cycle & 78.51 & 73.84 & 85.04 \\
    \hline
    \multicolumn{4}{l}{\textit{Ablating Specific Mechanisms}} \\
    \quad w/o Completeness Check & 80.93 & 71.62 & 86.15 \\
    \quad w/o Relevance Check & 81.65 & 72.88 & 86.55 \\
    \quad w/o Strategic Restart & 82.81 & 73.54 & 87.25 \\
    \hline
    \end{tabular}
\end{table}

\begin{figure*}[htbp]
  \centering
  \includegraphics[width=0.85\linewidth]{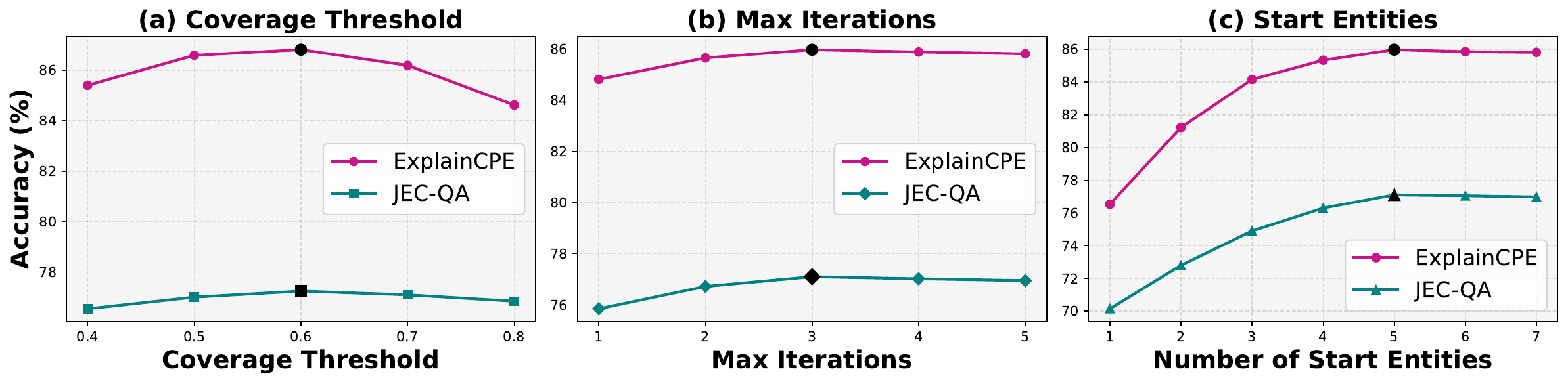}
  \caption{Evaluation of different hyperparameters on ExplainCPE and JEC-QA.}
  \label{fig:param-test}
\end{figure*}

\subsubsection{Detailed Analysis}
To gain deeper insights into the internal mechanisms and effectiveness of our MetaKGRAG framework, we conduct a series of detailed analyses across multiple dimensions.

\begin{figure}[htbp]
  \centering
  \includegraphics[width=0.85\linewidth]{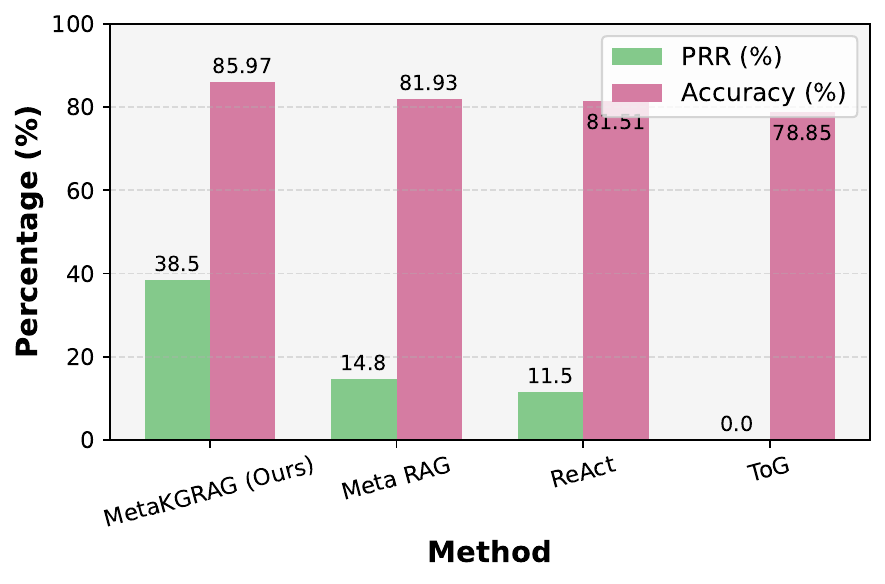}
  \caption{Path refinement rate (PPR) and accuracy of different methods on ExplainCPE.}
  \label{fig:retrieval_quality}
\end{figure}

\textbf{Effectiveness Analysis.} We first verify whether MetaKGRAG's metacognitive cycle genuinely improves evidence quality through effective path refinement. We introduce the Path Refinement Rate (PRR) metric to quantify the degree of change in final evidence paths relative to initial paths. As shown in Fig.~\ref{fig:retrieval_quality}, our MetaKGRAG method demonstrates a high path refinement rate of 38.5\% while achieving 85.97\% accuracy. In contrast, baselines that simply incorporate self-refinement methods (Meta RAG and ReAct) exhibit refinement rates below 15\% with correspondingly lower accuracy, while ToG shows zero refinement. This strong correlation between a high refinement rate and high accuracy proves that our adjustments are targeted and beneficial; our method makes more meaningful changes to improve answer quality, whereas other methods either do not refine the path or make fewer, less effective adjustments.

\textbf{Parameter Sensitivity Analysis.} We examine the framework's sensitivity to two core hyperparameters: the coverage threshold $\tau_{coverage}$ (used to judge path entity coverage of question key concepts) and the maximum iteration number $N_{max}$. As shown in Fig.~\ref{fig:param-test} (a), $\tau_{coverage}$ performs optimally around 0.6 across both datasets. This is because a threshold set too low risks triggering unnecessary adjustments for irrelevant concepts, while one set too high may fail to identify and correct important evidence gaps. For $N_{max}$, as shown in Fig.~\ref{fig:param-test} (b), its performance peaks at 3 iterations and then stabilizes, suggesting that most path deficiencies can be effectively resolved within three cycles, with further iterations offering diminishing returns. The consistency of these optimal values across different domains validates our default parameter choices and demonstrates the framework's robustness.

\textbf{Multi-start Retrieval Analysis.} To quantify the importance of the multi-start retrieval strategy, we analyze the impact of the number of starting entities on final performance. As illustrated in Fig.~\ref{fig:param-test} (c), accuracy on both datasets improves significantly as the number of starting entities increases, reaching an optimal point at 5 before plateauing. This is because complex questions often require multiple perspectives to construct a complete evidence subgraph, and starting from just one entity frequently misses critical information. Notably, medical questions (ExplainCPE) benefit more from additional starting points than legal questions (JEC-QA), likely because medical problems often involve more interconnected concepts that necessitate exploration from diverse angles. This result echoes our ablation study conclusions, further demonstrating that exploring knowledge graphs from multiple perspectives is crucial for ensuring retrieval comprehensiveness. \textbf{Please refer to Appendix~\ref{appx:exp-results} for more experimental results analysis and case study.}

\section{Related Work}
\label{relatedworks}
% To contextualize our work, we review three interconnected research areas, including Knowledge Graph RAG (KG-RAG), general Self-Refinement Methods, and Metacognition. Our review is structured to build a clear argument, while existing KG-RAG methods lack quality awareness and general self-refinement techniques are ill-suited for the unique challenges of graph exploration.

\textbf{Knowledge Graph RAG.} Recent advances in KG-RAG have demonstrated the value of structured knowledge for complex reasoning tasks. ToG (Think-on-Graph)~\cite{Sun2023ThinkonGraphDA} guides LLMs to explore multiple reasoning paths within knowledge graphs through a beam search mechanism, where the model iteratively selects the most promising entities and relations to expand the current path. MindMap~\cite{wen-etal-2024-mindmap} constructs hierarchical structured representations by first identifying key entities, then building subgraph clusters around them, and finally integrating these clusters into a unified knowledge structure for enhanced interpretability. KGGPT~\cite{kim-etal-2023-kg} handles complex queries through a decomposition strategy, breaking down multi-faceted questions into simpler sub-queries and constructing evidence graphs through separate retrieval processes for each component. Despite their sophisticated exploration strategies, these methods operate as open-loop systems that generate evidence paths in a single forward pass without mechanisms to assess or refine path quality during the retrieval process.

\textbf{Self-Refinement Methods.} The importance of quality control in retrieval has been recognized in the broader RAG domain, leading to several self-refinement approaches. ReAct~\cite{yao2023react} introduces a reasoning-acting loop where the system can perform new searches based on intermediate reasoning states, using action sequences like "Search[query]" and "Finish[answer]" to iteratively gather evidence. FLARE~\cite{jiang-etal-2023-active} employs a forward-looking approach that generates answers incrementally, triggering retrieval when confidence drops below a threshold, thus ensuring continuous evidence support throughout generation. Self-RAG~\cite{asai2024selfrag} implements a more comprehensive framework with reflection tokens that enable the model to critique its own outputs and decide whether to retrieve additional information, revise responses, or continue generation. These methods excel in document-based retrieval, where individual pieces of evidence can be independently assessed and substituted. However, this approach mismatches the path-dependent nature of graph exploration, where each step constrains subsequent choices and simple replacement invalidates entire reasoning chains.

\section*{Conclusion}
In this paper, we proposed MetaKGRAG, a novel framework that enhances Knowledge Graph RAG by integrating a human-inspired metacognitive process to solve Cognitive Blindness. Through its  Perceive-Evaluate-Adjust cycle, MetaKGRAG empowers the system to identify path-level deficiencies like incomplete evidence and relevance drift, and to perform targeted, trajectory-connected corrections. Experimental results across diverse medical, legal, and commonsense datasets demonstrated the superior performance of MetaKGRAG over strong baselines. For future work, we aim to explore more advanced, learnable adjustment strategies to further enhance the efficiency and adaptability of the metacognitive cycle.

%% The next two lines define the bibliography style to be used, and
%% the bibliography file.
\bibliographystyle{ACM-Reference-Format}
\bibliography{biblio}

% \clearpage

\appendix

\section{Complementary experimental settings}
\label{appx:exp}

\subsection{Datasets}
\label{appx:exp-data}

The detailed descriptions of the adopted datasets are summarized as follows:
\begin{itemize}[leftmargin=*]
  \item CommonsenseQA~\cite{talmor-etal-2019-commonsenseqa} is a multiple-choice QA dataset specifically designed to evaluate commonsense reasoning capabilities. 
  Each question is accompanied by five candidate answers, only one of which is correct.
  
  \item CMB-Exam~\cite{wang-etal-2024-cmb} covers 280,839 questions from six major medical professional qualification examinations, including physicians, nurses, medical technologists and pharmacists, as well as Undergraduate Disciplines Examinations and Graduate Entrance Examination in the medical field at China.
  Given the extensive scale of CMB-Exam, we sample a subset of CMB-Exam that comprises 2,000 questions, where 500 questions are randomly sampled from nurses, pharmacists, Undergraduate Disciplines Examination, and Graduate Entrance Examination categories.
  
  \item ExplainCPE~\cite{li-etal-2023-explaincpe} is a Chinese medical benchmark dataset containing over 7K instances from the National Licensed Pharmacist Examination. This dataset is distinctive in providing both multiple-choice answers and their corresponding explanations.

  \item webMedQA~\cite{he2019applying} is a large-scale Chinese medical QA dataset constructed from professional health consultation websites (such as Baidu Docto). The dataset contains 63,284 questions covering various clinical departments including internal medicine, surgery, gynecology, and pediatrics, with answers provided by doctors and experienced users. The dataset has been preprocessed to remove web tags, links, and garbled characters, retaining only Chinese and English characters, numbers, and punctuation.

  \item JEC-QA~\cite{zhong2019jec} is a legal domain dataset collected from the National Judicial Examination of China. It serves as a comprehensive evaluation of professional skills required for legal practitioners. The dataset is particularly challenging as it requires logical reasoning abilities to retrieve relevant materials and answer questions correctly.
\end{itemize}

\subsection{Implementation details}
\label{appx:exp-setting}
Our framework is built on LangChain\footnote{https://www.langchain.com/}. The local open-source LLMs are deployed based on the llama.cpp\footnote{https://github.com/ggml-org/llama.cpp} project. Except for the context window size, which is adjusted according to the dataset, all other parameters use default configurations, such as temperature is 0.8. Both LangChain and llama.cpp are open-source projects, providing good transparency and reproducibility. For computational resources, all experiments were conducted on a cluster of 8 NVIDIA RTX 3090 GPUs. Due to computational constraints and to ensure fair comparison across different model scales, we applied 4-bit quantization to locally deployed LLMs.
For the evaluation, we employed Bert Score metrics using ``bert-base-chinese~\cite{devlin-etal-2019-bert}'' model, while ROUGE Score version 0.1.2 was utilized. Due to resource constraints, G-Eval assessments were conducted using locally deployed Qwen2.5-72B.

\section{Knowledge Graph Construction}
\label{appx:KG-Construction}
We employed a consistent KG construction method for all datasets, utilizing LLMs to extract knowledge triples from the datasets to build specialized KGs. The prompt example is shown in Tab.~\ref{tab:prompt_kg_construction}. All KGs were deployed using Neo4j\footnote{https://neo4j.com/}.

\section{Complementary experimental results}
\label{appx:exp-results}
\subsection{Analysis of Generative Quality}
In addition to evaluating answer accuracy, we assessed the quality of the generated explanations for tasks requiring free-form responses (ExplainCPE and webMedQA). We used ROUGE-L to measure lexical overlap with reference answers and G-Eval to evaluate the Coherence, Consistency, Fluency, and Relevance of the generated text.
The results, presented in Table~\ref{tab:results_rouge_geval}, demonstrate that MetaKGRAG's advantages extend beyond correctness to significantly enhance generative quality. Across both datasets, MetaKGRAG consistently achieves the highest scores in all evaluated dimensions. For instance, on ExplainCPE, MetaKGRAG (Qwen2.5-72B) achieves a ROUGE-L score of 28.45, a substantial improvement of over 4.4 points compared to the best-performing baseline (Meta RAG).
This significant improvement can be attributed to the high-quality evidence subgraphs produced by our metacognitive cycle. The path-aware refinement process does not just retrieve relevant facts; it constructs a coherent, logically connected evidence narrative. This well-structured context enables the LLM to generate answers that are not only factually accurate but also more fluent, consistent, and directly relevant to the user's query. In contrast, while other retrieval methods provide a performance lift, the potentially fragmented or incomplete evidence they retrieve limits the ultimate quality of the generated text. This analysis confirms that the path-level self-correction of MetaKGRAG is crucial for both accuracy and the quality of explanatory generation.

\subsection{Analysis of Concept Relevance Threshold}
\label{appx:relevance-threshold}

The concept relevance threshold, $\tau_c$, is a hyperparameter within the Evaluation module used to determine if an entity provides meaningful support for a query concept (i.e., if $\text{sim}(e, c) > \tau_c$). A threshold set too low might accept noisy entities, while one set too high could prematurely discard useful ones. To assess the framework's sensitivity to this parameter, we conducted a tuning experiment on both ExplainCPE and JEC-QA datasets, varying $\tau_c$ from 0.1 to 0.5. The results are illustrated in Figure~\ref{fig:appendix_param_analysis} (a).
As shown in the figure, the model's performance remains highly stable across the entire range of $\tau_c$ values. For both datasets, the accuracy fluctuates by less than 0.4\%, with a slight peak around $\tau_c = 0.3$. This indicates that while a reasonably calibrated threshold is beneficial, MetaKGRAG is not overly sensitive to its precise value. The framework's robustness in this regard simplifies its deployment in new domains, as it does not require extensive, dataset-specific tuning of this parameter.

\subsection{Analysis of Weight Adjustment}
\label{appx:weight-adjustment}
\begin{figure}[htbp]
  \centering
  \includegraphics[width=0.95\linewidth]{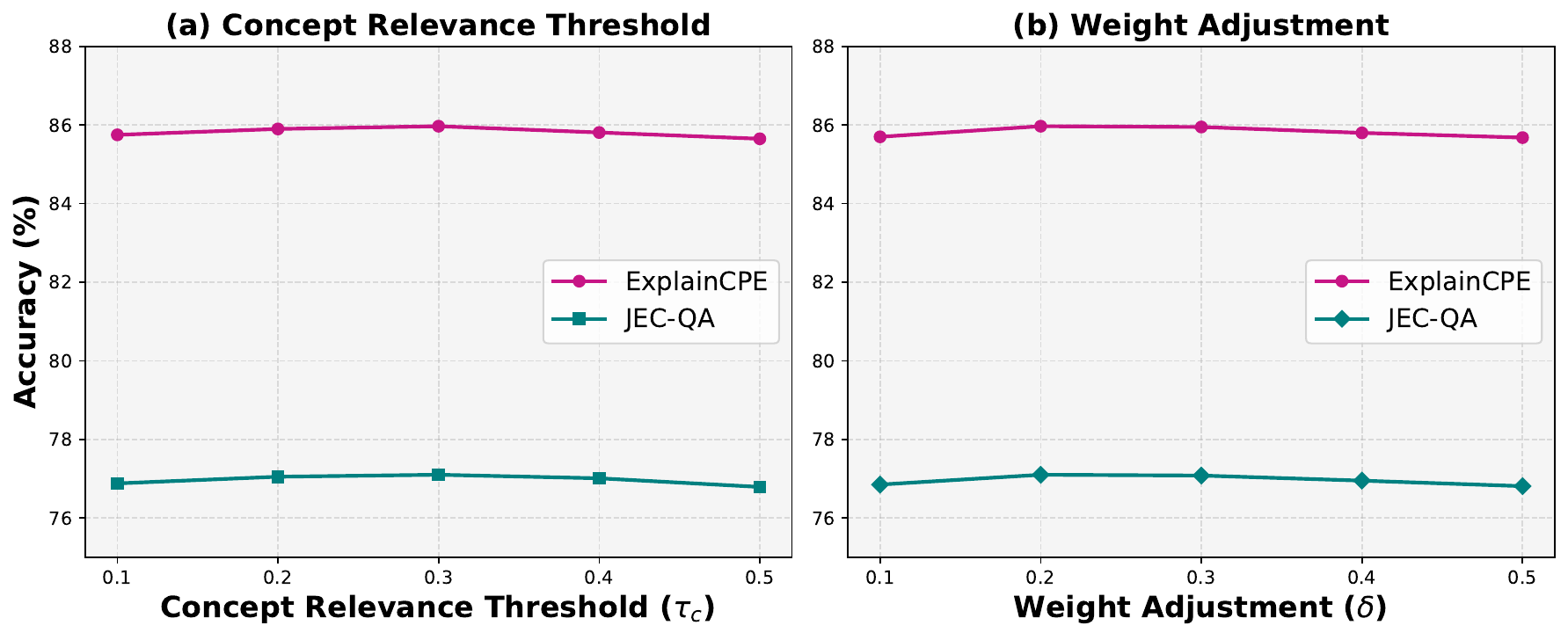}
  \caption{Evaluation of other hyperparameters on ExplainCPE and JEC-QA.}
  \label{fig:appendix_param_analysis}
\end{figure}

The weight adjustment parameter, $\delta$, controls the magnitude of the positive or negative incentive applied to entities during the Strategic Re-exploration phase of the Adjustment module. A larger $\delta$ more aggressively steers the search away from misleading entities and towards missing concepts. We validated the impact of $\delta$ by testing values from 0.1 to 0.5.
The results, depicted in Figure~\ref{fig:appendix_param_analysis} (b), demonstrate that the framework exhibits low sensitivity to the specific value of $\delta$. Performance on both datasets forms a plateau, with optimal results achieved for $\delta$ values between 0.2 and 0.3. The minimal variation in accuracy suggests that as long as the adjustment provides a clear directional signal, its exact magnitude is not a critical factor.

\begin{table*}[b] 
  \caption{Prompt Example for Answer Generation}
    \centering
    \label{app:answer_prompt}
    \begin{adjustbox}{max width=1\linewidth}  
        \begin{verbbox}
  prompt = f"""Your task is to accurately understand the question requirements and provide 
            reasonable answers and explanations based on the provided reference content.
            
            Input Question:
            {question}
            
            You have the following medical evidence knowledge:

            {evidence_text}
            
            What is the answer to this multiple-choice question? Answer the question by 
            referring to the provided medical evidence knowledge. First, choose the answer 
            from (A\B\C\D\E), output the answer option, then explain the reasoning.
            """
        \end{verbbox}
        \theverbbox
    \end{adjustbox}
  \end{table*}

\begin{table*}[b] 
  \caption{Prompt Example for Knowledge Graph Construction}
    \centering
    \label{tab:prompt_kg_construction}
    \begin{adjustbox}{max width=1\linewidth}  
        \begin{verbbox}
  prompt = f"""As a professional knowledge extraction assistant, your task is to extract knowledge triples from the given question.
  1. Carefully read the question description, all options, and the correct answer.
  2. Focus on the core concept "{question_concept}" in the question.
  3. Extract commonsense knowledge triples related to the question.
  4. Each triple should be in the format: subject\predicate\object
  5. Focus on the following types of relationships:
   - Conceptual relations 
   - Object properties 
   - Object functions 
   - Spatial relations 
   - Temporal relations 
   - Causal relations 
  6. Each triple must be concrete and valuable commonsense knowledge.
  7. Avoid subjective or controversial knowledge.
  8. Ensure triples are logically sound and align with common sense.
  
  Please extract knowledge triples from this multiple-choice question:
  
  Question: {question}
  Core Concept: {question_concept}
  Correct Answer: {correct_answer}
  
  Please output knowledge triples directly, one per line, in the format: subject\predicate\object. """
        \end{verbbox}
        \theverbbox
    \end{adjustbox}
  \end{table*}

\subsection{Case Study}
\label{appx:case-study}
We conduct a case study on the ExplainCPE, . 
To visually demonstrate the operational difference between MetaKGRAG and baseline methods when handling complex queries, we present a representative case from our medical dataset. As shown in Tab. ~\ref{tab:case_study}, this case illustrates the ``Cognitive Blindness'' issue in traditional KG-RAG, the path-dependency failure of self-refinement methods, and showcases how our metacognitive cycle effectively corrects the evidence-gathering path.

\begin{table*}[b]
\centering
\caption{Comparison of ROUGE-L and G-Eval scores on ExplainCPE and webMedQA. Coh., Cons., Flu., and Rel. indicate Coherence, Consistency, Fluency, and Relevance, respectively.}
\label{tab:results_rouge_geval}
\footnotesize
\setlength\tabcolsep{4.5pt}
\renewcommand{\arraystretch}{1.2}
\begin{tabular}{c|c|c|cccc|c|cccc}
\hline
\multicolumn{2}{c|}{} & \multicolumn{5}{c|}{\textbf{ExplainCPE (Medical)}} & \multicolumn{5}{c}{\textbf{webMedQA (Medical)}} \\
\cmidrule(lr){3-7} \cmidrule(lr){8-12}
\multicolumn{2}{c|}{\multirow{-2}{*}{\textbf{Method}}} & \textbf{ROUGE-L} & \textbf{Coh.} & \textbf{Cons.} & \textbf{Flu.} & \textbf{Rel.} & \textbf{ROUGE-L} & \textbf{Coh.} & \textbf{Cons.} & \textbf{Flu.} & \textbf{Rel.} \\
\hline
\multicolumn{12}{c}{\textbf{Without Retrieval}} \\
\hline
\multirow{6}{*}{LLM Only} & Qwen2.5-7B & 18.55 & 95.11 & 90.23 & 91.88 & 85.74 & 20.15 & 94.88 & 89.95 & 90.15 & 84.69 \\
& Qwen2.5-72B & 21.34 & 96.53 & 92.88 & 93.15 & 88.91 & 23.41 & 96.15 & 91.83 & 92.55 & 87.81 \\
& GPT4o & 20.89 & 96.81 & 93.05 & 93.55 & 89.15 & 22.95 & 96.53 & 92.11 & 92.98 & 88.05 \\
& o1-mini & 17.98 & 94.88 & 89.75 & 91.03 & 85.01 & 19.88 & 94.13 & 89.15 & 89.87 & 84.11 \\
& Claude3.5-Sonnet & 19.52 & 95.83 & 91.54 & 92.51 & 87.13 & 21.73 & 95.77 & 90.89 & 91.83 & 86.58 \\
& Gemini1.5-Pro & 20.15 & 96.11 & 92.18 & 92.88 & 88.05 & 22.51 & 96.01 & 91.55 & 92.71 & 87.92 \\
\hline
\multirow{2}{*}{Self-Refine} & CoT & 21.88 & 96.95 & 93.51 & 93.88 & 89.53 & 23.98 & 96.83 & 92.55 & 93.18 & 88.51 \\
& Meta Prompting & 22.15 & 97.03 & 93.88 & 94.01 & 89.98 & 24.33 & 97.01 & 92.93 & 93.55 & 88.99 \\
\hline
\multicolumn{12}{c}{\textbf{With Retrieval}} \\
\hline
\multirow{4}{*}{KG-RAG} & KGRAG & 22.58 & 97.15 & 94.01 & 94.18 & 90.15 & 24.88 & 97.18 & 93.11 & 93.88 & 89.53 \\
& ToG & 22.91 & 97.33 & 94.52 & 94.55 & 90.83 & 25.13 & 97.51 & 93.82 & 94.13 & 90.11 \\
& MindMap & 22.43 & 97.01 & 93.85 & 94.03 & 89.95 & 24.71 & 97.05 & 92.95 & 93.72 & 89.25 \\
& KGGPT & 22.83 & 97.28 & 94.41 & 94.48 & 90.71 & 25.05 & 97.43 & 93.71 & 94.01 & 89.98 \\
\hline
\multirow{4}{*}{Self-Refine} & FLARE & 23.15 & 97.51 & 94.88 & 94.83 & 91.15 & 25.53 & 97.83 & 94.18 & 94.51 & 90.83 \\
& ReAct & 23.88 & 97.98 & 95.53 & 95.11 & 92.01 & 26.15 & 98.15 & 94.98 & 95.03 & 91.95 \\
& Meta Prompting & 23.41 & 97.72 & 95.01 & 94.95 & 91.53 & 25.81 & 97.99 & 94.51 & 94.82 & 91.21 \\
& Meta RAG & 24.03 & 98.05 & 95.81 & 95.33 & 92.35 & 26.33 & 98.33 & 95.21 & 95.38 & 92.18 \\
\hline
\multirow{2}{*}{Ours} & MetaKGRAG (Qwen2.5-7B) & \underline{25.17} & \underline{98.53} & \underline{96.58} & \underline{96.01} & \underline{93.51} & \underline{27.55} & \underline{98.81} & \underline{96.03} & \underline{96.15} & \underline{93.11} \\
& MetaKGRAG (Qwen2.5-72B) & \textbf{28.45} & \textbf{99.11} & \textbf{98.03} & \textbf{97.88} & \textbf{96.15} & \textbf{30.18} & \textbf{99.25} & \textbf{98.15} & \textbf{98.01} & \textbf{96.53} \\
\hline
\end{tabular}
\end{table*}

\begin{table*}[h!]
\centering
\caption{A comparative analysis of a drug interaction case, showing how MetaKGRAG overcomes the limitations of KG-RAG baseline and self-refinement method.}
\label{tab:case_study}
\begin{tabular}{p{\dimexpr\textwidth-2\tabcolsep\relax}}
\toprule
\textbf{Input Question}: Which of the following statements about medications for Alzheimer's disease is \textbf{incorrect}? \\
\quad A. Cholinesterase inhibitors can increase the risk of stomach bleeding. \\
\quad B. Donepezil is suitable for co-administration with NSAIDs. \\
\quad C. Memantine's clearance can be affected by urinary pH. \\
\quad D. Galantamine's metabolism can be inhibited by ketoconazole. \\
\midrule
\textbf{Baseline Method (Vanilla KG-RAG)}: \\
\textit{Analysis}: Employs a greedy search. It strongly associates \texttt{Donepezil} with its primary function, \texttt{treating Alzheimer's}, and explores this path, overlooking the interaction with \texttt{NSAIDs}. \\
\textit{Initial Path}: \\
\quad 1. \texttt{Donepezil} $\xrightarrow{\text{treats}}$ \texttt{Alzheimer's Disease} \\
\quad 2. \texttt{Donepezil} $\xrightarrow{\text{is\_a}}$ \texttt{Cholinesterase Inhibitor} \\
\textit{Problem Diagnosis}: \\
\quad \textbf{Incomplete Evidence}: The path completely fails to retrieve the critical information about the adverse interaction between Cholinesterase Inhibitors and NSAIDs. \\
\textit{Generate Answer}: Based on the retrieved evidence, the model cannot identify the incorrect statement. It might wrongly conclude that statement (B) is plausible. \\
\midrule
\textbf{Self-Refinement Method}: \\
\textit{Analysis}: After generating an initial path like the baseline, the refinement mechanism recognizes the answer is insufficient as it doesn't address \texttt{NSAIDs}. \\
\textit{Corrective Action (Flawed)}: It triggers a \textbf{separate and isolated search} for the interaction between \texttt{Donepezil} and \texttt{NSAIDs}. This returns a new, isolated fact: \texttt{Cholinesterase Inhibitors + NSAIDs $\rightarrow$ increased risk of GI bleeding}. \\
\textit{Problem Diagnosis (Path Dependency Failure)}: The system now has two \textbf{disconnected evidence fragments}. It cannot integrate the new interaction fact with the original path. It fails to build a coherent reasoning chain explaining \textit{why} the interaction occurs (e.g., via increased gastric acid secretion), and thus struggles to synthesize a confident and well-grounded response. \\
\textit{Generate Answer}: The model might mention a risk but cannot provide a clear explanation, or it may be confused by the fragmented evidence and fail to definitively identify (B) as incorrect. \\
\midrule
\textbf{Our Method (MetaKGRAG)}: \\
\textit{1. Initial Path Generation}: Generates a similar preliminary path: \texttt{Donepezil} $\rightarrow$ \texttt{treats} $\rightarrow$ \texttt{Alzheimer's Disease}. \\
\textit{2. Perceive-Evaluate-Adjust Cycle}: \\
\quad \textbf{Perceive}: The framework assesses the path's coverage and detects that the relationship with \texttt{NSAIDs} is missing. \\
\quad \textbf{Evaluate}: It diagnoses a \textbf{Completeness Deficiency}, flagging \texttt{co\_administration\_with\_NSAIDs} as the missing concept ($\mathcal{C}_{\text{missing}}$). \\
\quad \textbf{Adjust}: Instead of a disconnected search, it performs a \textbf{trajectory-connected correction}. It selects \texttt{Cholinesterase Inhibitor} as a strategic restart point to explore its class-level properties. \\
\textit{Refined Path (Coherent and Connected)}: \\
\quad 1. \texttt{Donepezil} $\xrightarrow{\text{is\_a}}$ \texttt{Cholinesterase Inhibitor} \\
\quad 2. \texttt{Cholinesterase Inhibitor} $\xrightarrow{\text{increases}}$ \texttt{Gastric Acid Secretion} \\
\quad 3. \texttt{Gastric Acid Secretion} $\xrightarrow{\text{heightens\_risk\_with}}$ \texttt{NSAIDs} \\
\quad 4. \texttt{NSAIDs} $\xrightarrow{\text{can\_cause}}$ \texttt{Gastrointestinal Bleeding} \\
\textit{Generate Answer}: With the complete and connected evidence path, the model understands the full causal chain: Donepezil increases gastric acid, which heightens the risk of bleeding when combined with NSAIDs. It thus confidently and correctly identifies statement (B) as \textbf{incorrect}. \\
\bottomrule
\end{tabular}
\end{table*}

% \begin{table*}[h!]
% \centering
% \caption{Case Study between XXX and MetaKGRAG.}
% \begin{tabular}{p{\dimexpr\textwidth-2\tabcolsep\relax}}
% \toprule
% \textbf{Input}: Where do Florida Panthers play? (from WebQSP) \\
% \midrule
% \textbf{XXX}: \\
% \textit{Depth-1}: \quad Florida Panthers $\rightarrow$ sports.sports\_team.location $\rightarrow$ Sunrise, Florida Panthers $\rightarrow$ sports.sports\_team.venue $\rightarrow$ UnName\_Entity, ... \\
% \textit{Depth-2}: \quad Sunrise $\rightarrow$ sports.sports\_team\_location.teams $\rightarrow$ Florida Panthers, UnName\_Entity $\rightarrow$ sports.team\_venue\_relationship.venue $\rightarrow$ Miami Arena, ... \\ \\
% \textit{Answer}: \quad Sunrise \\
% \midrule
% \textbf{MetaKGRAG (Ours)}: \\
% Florida Panthers $\rightarrow$ sports.sports\_team.arena\_stadium $\rightarrow$ Miami Arena[Partially Relevant], Florida Panthers $\rightarrow$ sports.sports\_team.arena\_stadim $\rightarrow$ BB\&T Center[Fully Relevant]. \\ \\
% \textit{Answer}: Miami Arena; BB\&T Center [Utility:4] \\
% \bottomrule
% \end{tabular}
% \label{tab:case_study}
% \end{table*}

% Jielong: 这个部分最好不放
% \section{Limitations}
% MetaKGRAG's refinement process is guided by several key hyperparameters, such as the coverage threshold and the path similarity threshold. While our sensitivity analysis demonstrates robust performance around the chosen default values, the optimal settings might vary across different domains or datasets. Deploying MetaKGRAG in new applications may require careful tuning to achieve peak performance.

\end{document}